\documentclass{article}
\usepackage{arxiv}
\usepackage{amsmath}
\usepackage{float}
\usepackage[utf8]{inputenc}
\usepackage[T1]{fontenc}
\usepackage{hyperref}
\usepackage{url}
\usepackage{booktabs}
\usepackage{amsfonts}
\usepackage{nicefrac}
\usepackage{microtype}
\usepackage{lipsum}
\usepackage{graphicx}
\usepackage{natbib}
\usepackage{doi}

\title{Sustainable Data Management: Indefinite Static Data at Rest with Machine-Readable Printed Optical Data Sheets (MRPODS)}

\author{ \href{https://orcid.org/0009-0004-9682-5193}{\includegraphics[scale=0.06]{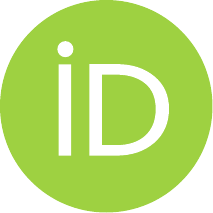}\hspace{1mm}Ray~Doll}\\
        Independent Researcher\\
	\texttt{sheafdynamics@gmail.com} \\
}

\renewcommand{\headeright}{}
\renewcommand{\undertitle}{}
\renewcommand{\shorttitle}{\textit{Sustainable Data Management: Indefinite Static Data at Rest with Machine-Readable Printed Optical Data Sheets (MRPODS)}}

\hypersetup{
pdftitle={Sustainable Data Management: Indefinite Static Data at Rest with Machine-Readable Printed Optical Data Sheets (MRPODS)},
pdfsubject={cs.CY, cs.DL},
pdfauthor={Ray~Doll},
pdfkeywords={Data Management, Data Archival, Machine-Readable Formats},
}

\begin{document}
\maketitle

\begin{abstract}
In an era where both commercial and private sectors place a premium on the longevity of digital data storage, the imperative to bolster resilience of digital information while simultaneously curbing costs and reducing failure rates becomes paramount. This study delves into the unique attributes of optical encoding methodologies, which are poised to offer enduring stability for digital data. Despite their promising potential, there remains a notable dearth of comprehensive analyses comparing various optical encoding techniques in terms of their durability. This research is thus dedicated to exploring the financial and environmental implications of employing technology to transcribe digital data into a machine-readable optical format, assessing both the advantages and limitations inherent in this approach. Our empirical findings reveal a marked increase in the efficiency of machine-readable optical encoding over conventional digital storage methods, particularly as the volume of data diminishes and the expected lifespan of storage extends indefinitely. This paper aims to illuminate key aspects of long-term digital data storage within business contexts, focusing on aspects such as cost, dependability, legibility, and confidentiality of optically encoded digital information.
\end{abstract}


\section{Introduction}
In contemporary business environments, documents containing sensitive information are essential for legal and competitive confidentiality. This category encompasses, but is not restricted to, confidential agreements (NDAs), trade secrets, legal contracts, financial records, client and customer data, intellectual property, employee records, strategic plans, research and development documents, corporate communications, IT infrastructure details, access codes, password databases, and asset valuations. These documents, while occupying minimal storage space, are perpetually at risk of cyber threats, including cyberattacks, digital corruption, ransomware (leading to data loss), and unauthorized alterations. Pre-emptively encrypted physical reproductions of such digital data present a less vulnerable alternative. In scenarios where digital infrastructures are compromised, the physical storage of critical information could be pivotal in maintaining operational continuity with minimal disruption.

Optical disks, traditionally considered robust, exhibit less durability than previously assumed, suffering from issues such as the degradation of their reflective layers, sensitivity to light exposure, and a lifespan falling short of two decades. Additionally, their storage capacity is comparatively limited, and they necessitate complex playback equipment reliant on motors and lasers, components with a propensity for failure and potentially limited manufacturing lifespan.

Conversely, printers and scanners, including basic consumer models, are expected to remain prevalent in the foreseeable technological landscape. Notably, while a CD/DVD reader inherently includes burning capabilities for data retrieval, paper scanners operate independently of printers. This independence, coupled with the ongoing evolution of digital storage technologies, necessitates the continued production of scanners for the foreseeable digitization of paper-based and printed texts. Such a scenario suggests that even if environmental considerations lead to the eventual obsolescence of printers within the next half-century to a century, previously printed data would remain accessible through scanning technology.

\section{Proof of Concept}
Oleh Yuschuk pioneered the printing of digital data in bitmap format onto paper, integrating Reed-Solomon error correction coding as a practical demonstration of concept \citep{yuschuk2007paperback}. Initially conceived with a semblance of programming humor, Yuschuk's innovation may not have fully captured the extensive potential of Machine-Readable Printed Optical Data Sheets (MRPODS). A notable advantage identified by Yuschuk is the visual verifiability of data integrity; the state of the data can be assessed without necessitating the activation of any electronic storage medium. This facet of physical encoding allows for the storage and assessment of data devoid of electricity or advanced technological interfaces, contingent upon the longevity and compatibility of the encoding and decoding software.

In an endeavor to ensure the future utility of his implementation, Yuschuk designed the software specifically for the 32-bit Windows platform, and elected to distribute it under the GPLv3 license. This licensing decision ensures the freedom to modify the source code, a critical consideration for future adaptability. Remarkably, the PaperBack software, despite its inception in 2007, remains functional on contemporary operating systems, including Windows 11 Pro 23H2 as of 2024, thereby encompassing nearly two decades of backward compatibility. This enduring functionality resonates particularly in business settings, where the integration and sustained operation of legacy software systems are prevalent.

However, should there be a paradigm shift in hardware away from x86 architecture without a corresponding compatibility framework, it is conceivable that physical encoding and decoding software might operate within a virtual machine environment, circumventing the need for direct source code alterations. Nonetheless, such a significant transition in hardware architecture would likely be accompanied by a prolonged adaptation phase, offering ample opportunity for entities to modify or port existing source code to align with future technological standards.

\section{Security Features Unique to MRPODS}
Data Sheets can be output at a sufficiently high dots per inch (DPI) resolution, effectively safeguarding them from unauthorized replication using standard equipment. In an era where smartphone cameras are ubiquitous, the risk of scanning and duplicating human-readable text is a pertinent concern. However, the fine granularity of physically encoded data in documents renders them impervious to replication by conventional camera technology. This attribute enables secure interdepartmental transfer of documents, mitigating the risk of data alteration or unauthorized copying during transit.

\begin{figure}[H]
    \centering
    \fbox{\includegraphics[width=10cm]{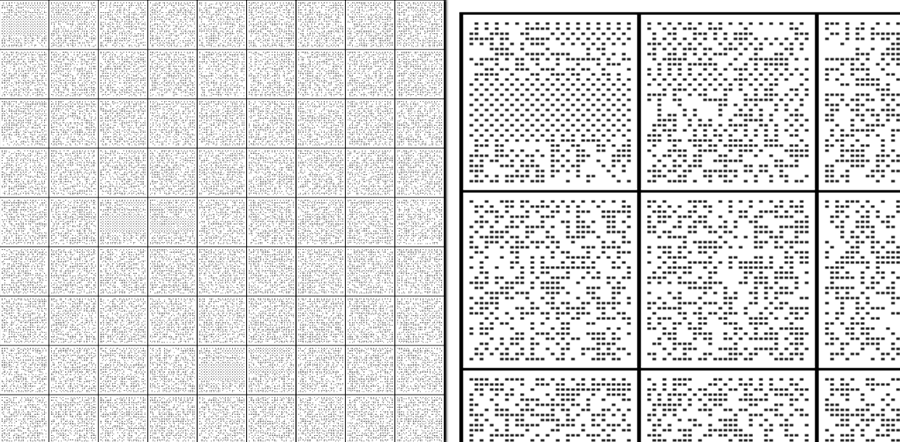}}
    \caption{Zoomed out view (left), and closer view (right)}
    \label{fig1:fig1}
\end{figure}

The robustness of documents against duplication can be further enhanced by meticulously adjusting various parameters such as the data DPI, dot size, and redundancy levels. Such optimization ensures that any attempt at scanning and reprinting these documents introduces an excessive amount of artifacts, thereby thwarting duplication efforts. Intrinsically, data sheets possess a layer of obfuscation; the bitmap data encoded in these documents is not decipherable by human cognition, providing a form of security even in the absence of additional encryption measures.

The structure of data sheets can be further tailored to include headers that detail the content or the file format, as well as the totality of pieces necessary for complete data retrieval. These headers are customizable and may deliberately omit information regarding the aggregate number of documents required for full data reconstruction until the commencement of the scanning process. This aspect serves as a strategic ambiguity, ensuring that individuals in possession of data sheets might remain oblivious to whether the document is encoded in its entirety, thereby augmenting security measures and deterring potential information theft.

\section{Data Density}

\section*{Data Compression Using \texttt{bzip2}}

The \texttt{bzip2} compression algorithm, as implemented in the \texttt{Printer.cpp} file, significantly increases data density through a combination of the Burrows-Wheeler Transform (BWT), the Move-to-Front (MTF) transform, and Huffman coding \citep{lauther2009space}.

Let $D$ be the original data stream. The BWT rearranges $D$ into $D'$, where similar characters are grouped together, thus forming runs of characters. This can be represented as:
\begin{equation}
D' = BWT(D)
\end{equation}

Subsequently, the MTF transform converts these runs into a more compressible form, denoted as $D''$:
\begin{equation}
D'' = MTF(D')
\end{equation}

Finally, Huffman coding is applied to $D''$, resulting in the compressed data stream $C$:
\begin{equation}
C = Huffman(D'')
\end{equation}

The overall compression process can be represented as:
\begin{equation}
C = Huffman(MTF(BWT(D)))
\end{equation}

In the context of data compression as implemented in \texttt{Printer.cpp}\footnote{The source code is available from: https://github.com/sheafdynamics/mrpods} of MRPODS using the \texttt{bzip2} algorithm, the compression ratio (CR) is a crucial metric. This ratio is defined as:

\begin{equation}
CR = \frac{ODS}{CDS}
\end{equation}

where \( ODS \) is the Original Data Size and \( CDS \) is the Compressed Data Size. The effectiveness of \texttt{bzip2} compression is measured by the increase in data density (DD), which is inversely proportional to the compressed data size. Thus, the data density can be expressed as:

\begin{equation}
DD = \frac{1}{CDS}
\end{equation}

This mathematical representation displays the efficiency of using \texttt{bzip2} in the MRPODS application. In a practical demonstration using a widely available consumer-grade HP OfficeJet 3830 printer, the efficacy of \texttt{bzip2} was evident when printing a WinRar compressed archive of the Project Gutenberg eBook version of 'War and Peace' (excluding the JPEG cover) on a mere five sheets of paper. The specific settings employed in this process were as follows: a dot density set at 200 dots per inch (dpi), dot size adjusted to 70\%, and a redundancy ratio of 1:5. The printing was executed in high DPI grayscale mode with the borderless printing option activated.

\begin{figure}[H]
    \centering
    \fbox{\includegraphics[width=15cm]{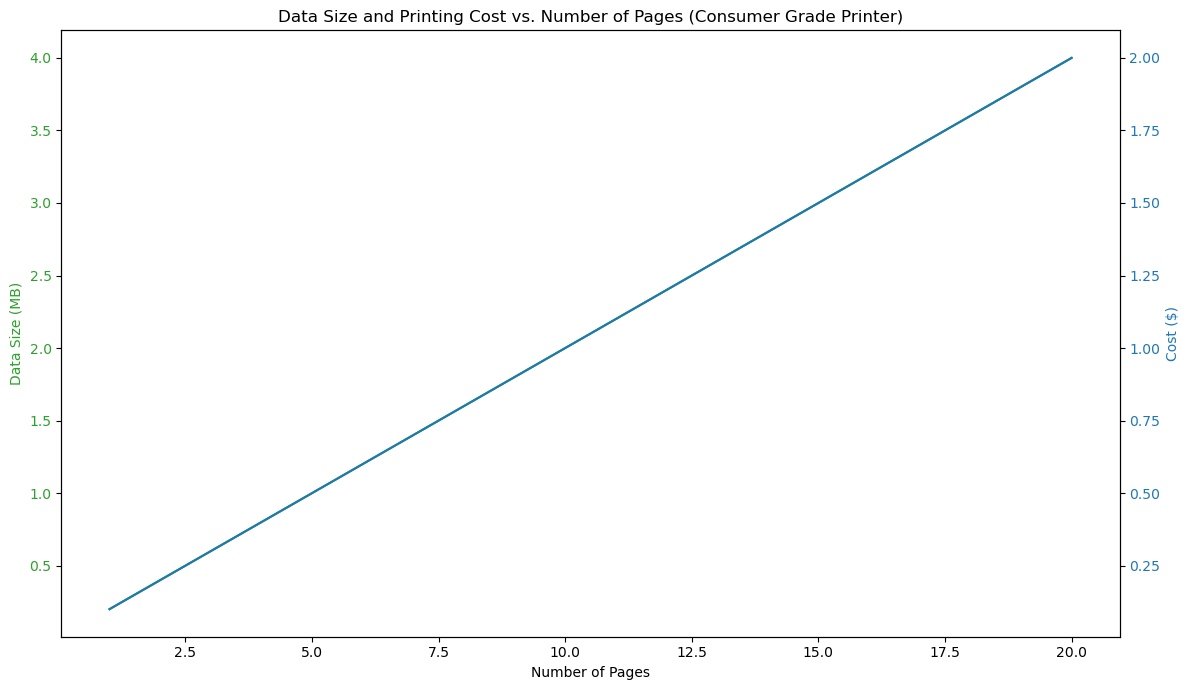}}
    \caption{How data density affects page output}\footnotemark
    \label{fig2:fig2}
\end{figure}

\footnotetext{This chart serves as an analytical tool, offering insights into the relationship between data volume and the economic aspects of MRPODS and the trade-offs involved.}

For comparison, when the same eBook was formatted for printing via Microsoft Word, utilizing a standard Times New Roman font at a 12-point size and narrow margins, the document spanned 748 pages. This stark contrast in page count between the two methods highlights the significant efficiency and space-saving potential of MRPODS for the storage of substantial textual data, provided the data does not exceed 4MB.

\section{Sustainability}

\begin{figure}[H]
    \centering
    \fbox{\includegraphics[width=15cm]{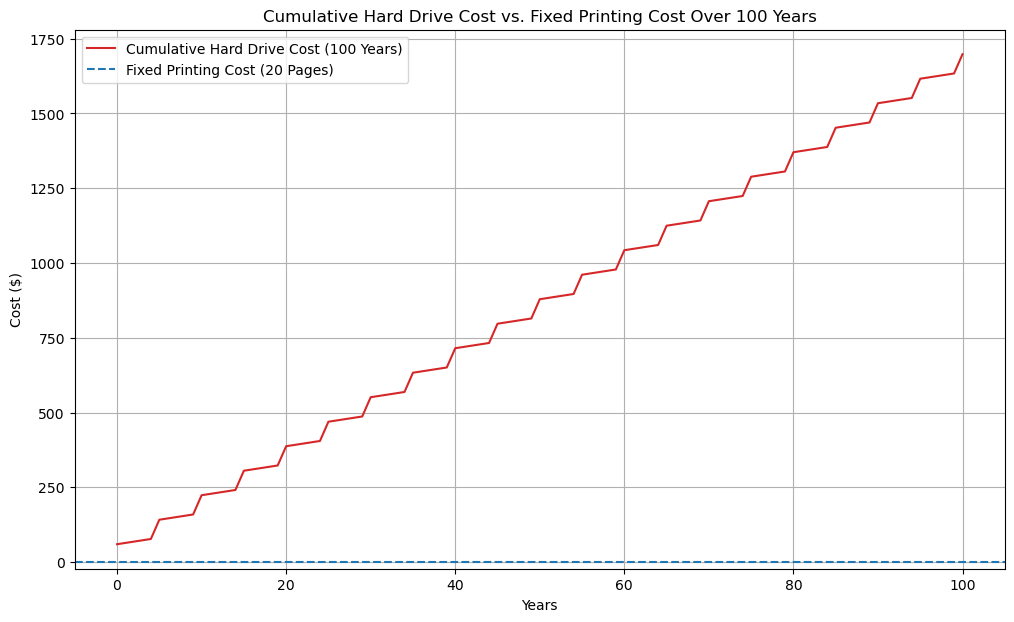}}
    \caption{Traditional storage vs. Data sheets cost}\footnotemark
    \label{fig3:fig3}
\end{figure}

\footnotetext{Visualized is the long-term fiscal and environmental sustainability of data management practices.}

\section{Conclusion}
In summary, the research presented in this paper shows possible advantages of Machine-Readable Printed Optical Data Sheets (MRPODS) in the context of secure and sustainable business data management. The data sheet paradigm emerges as a highly viable alternative to traditional digital storage methods, particularly in scenarios demanding long-term preservation and security of sensitive data. This approach effectively addresses key concerns such as data longevity, resistance to cyber threats, and the overarching issue of environmental sustainability.

The comparative analysis of MRPODS against conventional storage mediums displays its superiority in various critical aspects. Data sheets safeguard data against unauthorized access and duplication by using high-resolution optical encoding which resists standard replication techniques. The physical nature of data sheets inherently reduces the susceptibility of data to cyber threats, thus offering a more secure storage medium for sensitive and confidential business documents.

In terms of sustainability, the findings of this study reveal that MRPODS offers a more economically viable solution over an extended timeframe. The static cost associated with data sheets printing contrasts sharply with the escalating expenses of digital storage, which include recurrent hardware replacements and ongoing electricity costs. This attribute of MRPODS makes it particularly attractive for long-term data archiving, where cost-effectiveness is as crucial as data integrity.

The use of bzip2 within the MRPODS framework significantly enhances data density, allowing for the efficient storage of large volumes of information on minimal physical space. This aspect not only saves physical resources but also aligns with the broader environmental goals of reducing electronic waste and energy consumption.

Data sheets designed to be read by machines are a formidable contender in the realm of data storage, particularly for businesses seeking secure, sustainable, and cost-effective methods for managing their critical data. Machine-readable optical formats balance security, durability, and environmental responsibility. As businesses continue to navigate the complexities of sustainable data management in an increasingly digital world, data sheets stand out as a potential solution.

\bibliographystyle{unsrtnat}
\bibliography{references}

\end{document}